\documentclass{sigchi}

\CopyrightYear{2020}
\setcopyright{acmlicensed}
\doi{https://doi.org/10.1145/3313831.XXXXXXX}
\isbn{978-1-4503-6708-0/20/04}
\conferenceinfo{CHI'20,}{April  25--30, 2020, Honolulu, HI, USA}
\acmPrice{\$15.00}

\toappear{\scriptsize Permission to make digital or hard copies of all or part of this work for personal or classroom use is granted without fee provided that copies are not made or distributed for profit or commercial advantage and that copies bear this notice and the full citation on the first page. Copyrights for components of this work owned by others than ACM must be honored. Abstracting with credit is permitted. To copy otherwise, or republish, to post on servers or to redistribute to lists, requires prior specific permission and/or a fee. Request permissions from permissions@acm.org. \\
{\emph{CHI `20, April 25--30, 2020, Honolulu, HI, USA.} } \\
\copyright~2020 Association of Computing Machinery. \\
ACM ISBN 978-1-4503-6708-0/20/04\ ...\$15.00. \\
http://dx.doi.org/10.1145/3313831.3376746}
\usepackage{balance}       
\usepackage{graphics}      
\usepackage[T1]{fontenc}   
\usepackage{txfonts}
\usepackage{mathptmx}
\usepackage[pdflang={en-US},pdftex,pdfpagelabels=false]{hyperref}
\usepackage{color,soul} 
\usepackage{multirow}
\usepackage[table, dvipsnames]{xcolor}
\usepackage{booktabs} 
\usepackage{textcomp}
\usepackage{enumitem}
\usepackage{pifont}
\usepackage{array}
\usepackage{tabularx}
\usepackage{cite}
\usepackage{layouts}
\usepackage[caption=false]{subfig}

\usepackage{microtype}        
\usepackage{ccicons}          
\usepackage{sparklines}

\usepackage{url}

\usepackage{amsfonts}
\definecolor{brickred}{rgb}{0.8, 0.1, 0.1}
\definecolor{pastelyellow}{rgb}{0.99, 0.99, 0.59}
\usepackage{soul}
\sethlcolor{pastelyellow}

\usepackage{fancybox} 
\usepackage{tikz}
\def\techname{Spinneret}

\def\visualtitle{\techname: Aiding Creative Ideation through\\Non-Obvious Concept Associations\vspace{-0.5em}}

\def\plaintitle{\techname: Aiding Creative Ideation through Non-Obvious Concept Associations}

\def\emptyauthor{}

\def\plainkeywords{Mind mapping, Creativity, Knowledge Graph, Suggestion}

\makeatletter
\def\url@leostyle{%
  \@ifundefined{selectfont}{
    \def\UrlFont{\sf}
  }{
    \def\UrlFont{\small\bf\ttfamily}
  }}
\makeatother
\def\pprw{8.5in}
\def\pprh{11in}

\definecolor{linkColor}{RGB}{6,125,233}
\hypersetup{%
  pdftitle={\plaintitle},
  pdfauthor={\emptyauthor}, 
  pdfkeywords={\plainkeywords},
  pdfdisplaydoctitle=true, 
  bookmarksnumbered,
  pdfstartview={FitH},
  colorlinks,
  citecolor=black,
  filecolor=black,
  linkcolor=black,
  urlcolor=black,
  breaklinks=true,
  hypertexnames=false
}

\definecolor{sparkspikecolor}{rgb}{0,0,0}
\begin{document}

\title{\visualtitle}

\numberofauthors{1}
\author{%
  \alignauthor{Suyun ``Sandra'' Bae,\!$^\ast$
               Oh-Hyun Kwon,\!$^\ast$
               Senthil Chandrasegaran,
               and Kwan-Liu Ma\\
    \affaddr{Department of Computer Science,
             University of California, Davis}\\
    \email{\{suybae, kw, schandrasegaran, klma\}@ucdavis.edu}\\
    \vspace{-0.2em}\textbf{\small $^\ast$Equally-contributing authors}\vspace{-1.25em}
    }\\
}

\maketitle

\begin{abstract}
Mind mapping is a popular way to explore a design space in creative thinking exercises, allowing users to form associations between concepts.
Yet, most existing digital tools for mind mapping focus on authoring and organization, with little support for addressing the challenges of mind mapping such as stagnation and design fixation.
We present \techname, a functional approach to aid mind mapping by providing suggestions based on a knowledge graph.
\techname\ uses biased random walks to explore the knowledge graph in the neighborhood of an existing concept node in the mind map, and provides ``suggestions'' for the user to add to the mind map.
A comparative study with a baseline mind-mapping tool reveals that participants created more diverse and distinct concepts with \techname, and reported that the suggestions inspired them to think of ideas they would otherwise not have explored.
\end{abstract}


\vspace{-0.2em}
\vspace{-0.2em}

\section{Introduction}
Mind mapping is widely accepted as a tool for comprehension, reflective thinking, as well as creativity~\cite{Kokotovich2008problem}.
In creative problem solving, mind mapping is a way to explore divergent thinking, i.e., generating multiple solutions to address the problem.
Creative thinking can be seen as the ability to convert ``associative elements into new combinations by providing mediating connective links''~\cite[p.\ 226]{Mednick1962}.
Studies of ideation~\cite{Kudrowitz2013does} and mind-mapping outcomes for creative tasks~\cite{Leeds2019mapping} have shown that in order to come up with unique ideas (or nodes in a mind map), one needs to generate numerous ideas.
This is due to ``functional fixedness''~\cite{Flavell1958effect}---the mind's tendency to adhere to a fixed pattern of thinking---which restricts the creator from generating novel ideas at the beginning of the task.

While there have been procedural solutions to push the mind to unconventional thought processes, such as group brainstorming~\cite{Brown2002making}, 6-3-5 brainwriting~\cite{Linsey2011effectiveness}, and C-sketch~\cite{Shah2001collaborative}, most such solutions leverage group work as a way to mitigate the individual's fixation.
In addition, these processes are usually introduced in the context of sketching and brainstorming aspects of creative ideation, but not in the context of mind mapping.
While mind mapping as a team is possible, to the best of our knowledge, no team-oriented process has been proposed to address fixation in mind mapping.
In addition, few computer support tools have gone beyond extending these processes to the digital realm~\cite{Zhao2014skwiki, Clayphan2011firestorm, Hailpern2007team}.
Most existing digital tools for mind mapping~\cite{website:Mindomo, website:MindNode, website:MindMeister} focus on the physical aspects of the process, i.e.\ supporting the creation and reorganization of nodes and links, but do not actively help mitigate fixation.

We present \techname, a mind-mapping tool that aids creative thinking by providing non-obvious suggestions based on existing user-created concepts.
\techname\ aims to mimic the process of making non-obvious associations by exploring a knowledge graph in the neighborhood of a given
concept node through a biased random walk.
We explore two biases for the random walk: a breadth-first search bias, where the walk is more likely to stay within the immediate neighborhood of the source concept node, and a depth-first bias where the walk is less likely to stay within the immediate neighborhood of the source concept.
We implement \techname\ as a web-based interactive application 
for users to manually create nodes and links as well as request suggestions when they need inspiration.

We evaluate {\techname} through a controlled study, comparing the tool with a baseline mind-map authoring interface with no suggestion feature for two tasks: an open-ended task and a constrained task.
We find that while there is no significant difference in the number of nodes in mind maps created with either tool, mind maps created with \techname\ had more diverse concept nodes.
In addition, nodes created with \techname\ were more unique or distinct when compared to nodes created using the baseline.
In addition, we find that \techname's suggestions were better accepted during the open-ended task, while participants often did not find the suggestions from the constrained task useful regardless of the parameters controlling the biased random walk.
We conclude with suggestions to explore the use of word embeddings for a greater control for relevance based on the mind-mapping context.
\vspace{-2.5mm}

\section{Background}

Creative thinking can be described as the integration of associations that are novel and relevant to the problem at hand~\cite{Freedman1965increasing, Maltzman1960training}.
One of the primary barriers to creating novel associations is ``functional fixedness''~\cite{Flavell1958effect}: the inability to think of an artifact or a concept beyond one context.
Our goal in this work is to develop a mind-mapping tool that will help the user explore approaches and solutions that they would usually not consider.
In order to motivate our approach and understand our contributions, it is necessary to understand the aspects of mind maps that aid creative outcomes, understand the challenges posed by fixation, and explore computational approaches to mitigate fixation within and outside the context of mind mapping.

\subsection{Mind Mapping in Creative Tasks}

The notion of mind-mapping for idea generation was formally introduced
by Tony Buzan~\cite{Buzan1974use}.
He and others later argue that mind maps can be an inexhaustible source of ideas, stating, ``\ldots every key word or image added to a Mind Map itself adds the possiblity of a new and greater range of associations\ldots and so on \textit{ad infinitum}''~\cite[p.\ 86]{Buzan1993mind}.
Mind maps have the advantage of providing a simple and flexible view of
related concepts, while also allowing for structured thinking in terms of ``branches'' or categories when needed.
Drawing mind maps aids thinking as the physical representation keeps the ideas and their connections in what Ullman et al.~\cite{Ullman1988model} call ``the design state'': the combination of short- and long-term memories and external memory that helps designers solve complex problems.

While distinct from concept maps---top-down diagrams showing and explaining relationships between concepts~\cite{Novak1998learning}---mind maps are closely related to them when applied to tasks, such as learning and explaining.
Studies have shown that concept maps reveal the way their creators think and organize their thoughts~\cite{Ruiz2000use}.
In the same way, mind maps indicate the pattern and creativity of its creator.
Leeds et al.~\cite{Leeds2019mapping} show that the ``depth'' of a mind map---the average distance of a node from the central root node---is a predictor of creative output, the quantity of nodes correlate positively with the novelty of the concept nodes, and that nodes created later or deeper tend to be the more unique.
A study of mind map use in early-stage design~\cite{Kokotovich2008problem} showed that adding non-hierarchical links rendered to a mind map provides a better comprehension of
the problems, helps the creator develop their knowledge about their
problem, and serves as a better memory aid for reviewing problems.
A system that aids the creation of more nodes, the creation of links between nodes without adhering to a strict hierarchy, and the pursuit of ``deeper'' branches in the mind map can potentially aid a user's creative output in mind mapping.

\subsection{Fixation in Creative Work}
Functional fixedness is described as a (mind) set that makes it difficult to use an artifact in a way that is different from the habitual way to which one is accustomed through prior experience~\cite{Duncker1945problem, Flavell1958effect}.
The notion was initially coined as \textit{``Einstellung''} by Luchins~\cite{Luchins1942mechanization} after experiments that predisposed participants to solve a problem in a certain way, observing that the predisposition continued even when new problems that required different approaches were presented.
A similar effect, termed ``design fixation'' was observed by Jansson \& Smith~\cite{Jansson1991design} when they observed that when working on a design problem, subjects who were exposed to a solution sample tended to make very similar designs compared to subjects who were not shown a sample.

Creativity measures such as Guilford's Alternative Uses Test~\cite{Guilford1956structure} measure the lack of functional fixedness by asking subjects to list non-obvious uses for a given artifact, measuring fluency (number of uses), originality (uncommon uses), flexibility (different categories of uses), and elaboration (detail in the description of use).
Exercises such as free association thinking~\cite{Freedman1965increasing}---generating multiple spontaneous responses for the same stimulus word---are designed to overcome fixation and increase creative output.
Other methods to improve the quality of ideas include group ideation techniques where people can contribute to each other's ideas and negate individual biases.
Such methods include brainstorming~\cite{Osborn1953applied}, 6-3-5 brainwriting~\cite{Linsey2011effectiveness}, and C-Sketch~\cite{Shah2001collaborative}.
Though the three listed methods operate in different media (speech, writing, and sketching), the mechanism is the same: a group builds on each other's ideas to create novel combinations of concepts.
This idea of innovation by recombination has been studied at a larger scale by LaToza et al.~\cite{Latoza2015borrowing}, who found that borrowing ideas for software design from others in crowdsourced scenarios most improved participants' designs.
Even if not recombining, simply looking at creative and diverse ideas have been shown to influence the creativity and diversity of ideas generated~\cite{Siangliulue2015toward}.

Other domain-specific techniques include the morphological matrix~\cite{Zwicky1967morphological} in engineering design where a product's functions are broken down into sub-functions, and multiple ideas generated to achieve each sub-function.
The product design space then becomes a combinatorial explosion of the subfunction solutions.
More recently, Kudrowitz and Dippo~\cite{Kudrowitz2013does} used the Alternative Uses Test to show that most people's initial responses for the task end up being fairly common, with more original ideas occurring later in participant responses.
Calling this the ``long tail of originality'', they suggest keeping idea descriptions short in the beginning and focusing on the quantity of ideas to get to the more original ideas quickly.

\begin{figure*}[t]
    \captionsetup{farskip=0pt,skip=0pt}
    \centering
    \includegraphics[width=\textwidth]{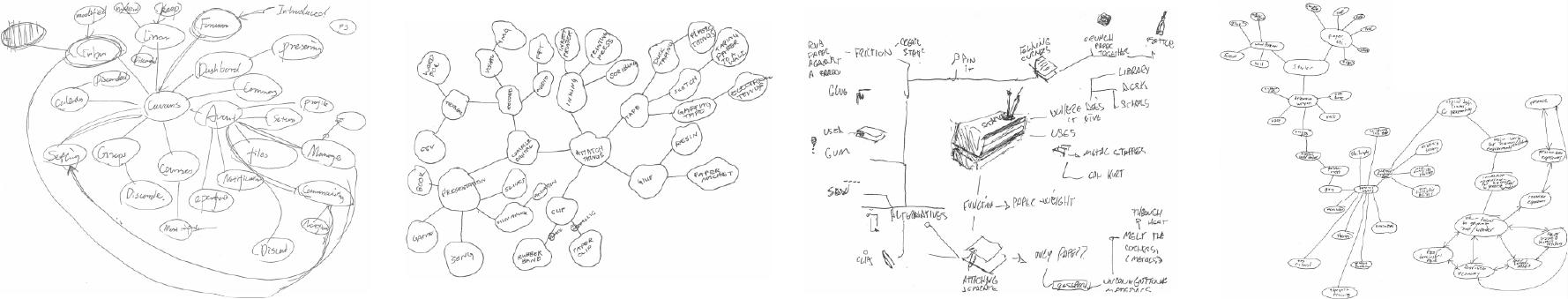}
    \caption{Samples from the preliminary study showing the diversity in the mind maps across participants.}
    \label{fig:prelim}
    \vspace{-1em}
\end{figure*}

\subsection{Computer Support for Creative Work}

In the prior section, we discussed several group ideation techniques such as brainstorming, 6-3-5 brainwriting, and C-Sketch that focused on expanding the individual's potential solution space.
Computational support tools for ideation often operationalize these group ideation techniques, while also taking advantage of the digital medium for scalablity, persistence, and easy duplication.
Some instances of digital support for group ideation include a brainstorming application that imposes best practices as constraints~\cite{Clayphan2011firestorm}, collaborative and mixed-media sketching frameworks~\cite{Hailpern2007team, Geyer2012ideavis, Zhao2014skwiki}, and pictorial stimulation for brainstorming based on verbal inputs~\cite{Wang2010idea, Shi2017ideawall}.

Computational support that aids the design search space includes search-based as well as suggestion-based approaches to either expand the search space, or suggest ``good'' design solutions.
In the context of engineering, good design typically translates to a combination of designs that work.
One approach has been used to automate the population of a morphological matrix using a repository of existing designs that would then increase the combinatorial space of potential designs for a new problem~\cite{Bohm2008using}.
A 3D shape-oriented approach to the same matrix is afforded by Co-3Deator~\cite{Piya2017co}, where the matrix is populated by the use of a component hierarchy at the conceptual stage of design, allowing the creation of modular components through collaborative modeling.
Juxtapoze~\cite{Benjamin2014juxtapoze} is a less constrained approach applied to clipart composition, supporting serendipity and creative exploration with a shape-based search.
Crowd-based approaches~\cite{Andolina2017crowdboard, Girotto2019crowdmuse, Girotto2017effect} incorporate ideation and feedback at large scale into specific parts or tasks in design problems.
Recent machine learning and knowledge-based approaches include using topic modeling to cluster ideas based on diversity, quality, and representation, from repositories maintained by open innovation communities~\cite{Ahmed2016discovering}.

While mind mapping as a technique is widely accepted for creative exploration, there is little computational support currently available for mind mapping.
Early tools such as gIBIS~\cite{Conklin1987gibis} and GENI~\cite{Maccrimmon1994stimulating} focused on characterizing connections and generation of alternatives.
To the best of our knowledge, existing digital tools~\cite{website:Mindomo, website:MindNode, website:MindMeister} typically focus on the mechanical and cosmetic aspects of mind mapping, such as (manually) creating and reorganizing nodes, and controlling color and layout of nodes and links.
The closest to our proposed work is a mixed-initiative mind-mapping tool by Chen et al.~\cite{Chen2019mini}.
They use the same knowledge graph (ConceptNet) as we do, but their approach is different from ours in two major ways: they take a concept map-like approach for the initial suggestions with the nodes representing \emph{relationships} and not concepts from ConceptNet, and their concept suggestions are created from the immediate neighborhoods of existing nodes.
In contrast, we use a biased random walk with a breadth-first or depth-first bias, suggest nodes typically not in the immediate neighborhood of the source concept, and do not suggest relationships between concepts.

\section{Preliminary Study}

In order to get a sense of the mind-mapping process and attributes of the mind map, we conducted a preliminary study of six design
students performing two creative mind-mapping tasks on paper.
We chose design students as they are typically trained in creative techniques including mind mapping, and 
to reflect on their processes.

\subsection{Study Setup}

The participants (4 female, 2 male, aged 18--60 years) were master's students in design, with backgrounds in digital art and design, architecture, human-computer interaction, and fashion and costume design.
All participants had prior experience in creating mind maps: 3 reported having created them on fewer than 10 occasions, 2 on 20--50 occasions, and one on more than 50 occasions.
All participants reported having used mind maps for both creative (e.g.,\ brainstorming) and analytical (e.g.,\ SWOT analysis, project planning) applications.
We gave each participant two mind-mapping tasks (10 minutes each), both involving creative thinking.
One was an unconstrained design task to find alternative ways to achieve the same function that a stapler does.
The second was a redesign task, and thus more constrained. Participants could choose between two redesign tasks: a backpack redesign, or a redesign of the university's course webpage.
The choice allowed participants to choose an option that was closer to their domain of comfort.
They were also allowed to use the internet to refer current designs to inform their redesign approach.
No external references were allowed for the unconstrained design task.

We used a concurrent think-aloud protocol for both tasks, and participants were audio- and video-recorded.
At the end of each tasks, participants were asked to explain the organization of their mind map, and answer a set of open-ended questions where they reflected on the task.
We also analyzed their mind maps to identify structural and thematic patterns.
\vspace*{-5pt}

\subsection{Observations}
\textbf{Structure.}
While the mind maps created were predominantly hierarchical, we found that four participants (2 novices, 2 experienced in mind mapping) created non-hierarchical links between existing nodes in the mind map.
Non-hierarchical connections occurred in 4 (out of 6) mind maps for the unconstrained task, and in 2 mind maps for the constrained task.

\textbf{Organization.}
Two of the participants (1 novice, 1 experienced) used text from the redesign task prompt to create organizing themes for their mind maps.
Of the remaining participants, there was no overt categorical organization, but on reflection, two participants pointed to thematic grouping in their mind maps.
For instance, one participant had a spatial organization where nodes above the root node were ``positives'' and the ones below were ``negatives'', referring to features of the backpack in the redesign task.
One participant neither organized their mind maps as categories, nor adhered to the design prompt for either task.
Instead, his mind map represented a free-association thinking starting from the design prompt and ending in themes that he found relevant to himself.

\textbf{Nodes.}
For 5 of the 6 participants, the root node was central to the mind map, with all other nodes radiating outward from the root or its children.
One participant created a single node (``sling'') from the root node (``backpack''), after which all subsequent nodes radiated outward from the ``sling'' node.
This also represented his thinking, which focused on the sling rather than the backpack for the remainder of the task.

\textbf{Links.}
Three participants created mind maps with non-directional links.
Of the remaining 3, one participant used arrows to emphasize hierarchy, while another used arrows to emphasize a non-hierarchical link.
The third participant used bidirectional links on one region of his mind map, but did not explain why.
None of the participants attempted to describe the nature of any links, nor did the links represent the same kind of relationship within any mind map.

\textbf{Contiguity.}
For both tasks, 4 of the 6 participants created a mind map on a single sheet, while 2 used multiple sheets.
Of these, one participant created a contiguous mind map across multiple sheets stuck together, while the other created two separate but related mind maps.

\textbf{Representation.}
All the mind maps were predominantly text-based, with 2 participants using sketches sparingly (see Fig.~\ref{fig:prelim} for an example).
For one participant, the sketch was an external representation of his memory of a stapler, and helped him think of its form, context of use, and features to add (he later added a ``pen holder'' feature to the sketch, saying the sketch helped him think of the idea).
The second participant used a sketch when they could not recall what a brass paper fastener was called, but remembered its appearance.

\vspace*{-3pt}
\section{Requirements and Design}

The design space of mind-mapping tools is fairly diverse, with most providing an editing function only, where the tool combines paper-based mind mapping with the flexibility of a digital canvas.
However, our goal is to develop an ideation tool that uses the practice of mind-mapping to help users get to the ``long tail of originality''~\cite[p.\ 16]{Kudrowitz2013does}, i.e.\ ideas that are more distinct from those that usually occur to designers.

\vspace*{-3pt}
\subsection{Requirements}
Based on existing literature and on our preliminary mind-mapping study, we identify a set of design requirements.

\begin{figure*}[t]
    \centering
    \includegraphics[width=\textwidth]{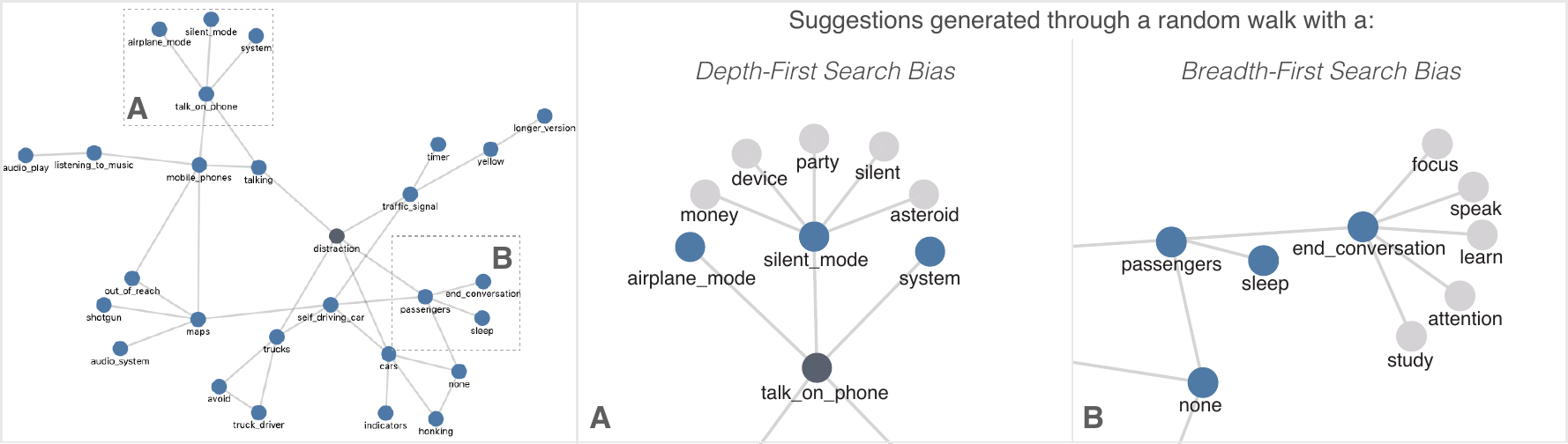}
    \caption{A mind map created with {\techname} (left).
    \techname\ provides node suggestions (grey nodes) to the user using random walks on a knowledge graph in the neighborhood of a selected node.
    The random walks can have (A) a depth-first search bias, generating suggestions from concepts farther away from the source node, or (B) a breadth-first search  bias such that suggestions are generated from concepts closer to the source node.}
    \label{fig:interface}
    \vspace{-1em}
\end{figure*}

\begin{enumerate}[itemindent=1.9em,leftmargin=0em,itemsep=0em]

  \item[\textbf{R1.}] \textbf{Reduce Design Fixation:}
    When approaching design as a problem-solving exercise, designers often exhibit an inhibition that prevents them from searching outside the known space of existing designs.
    This inhibition, termed ``design fixation''~\cite{Jansson1991design} hinders creativity.
    A mind-mapping tool for creative problem solving should reduce design fixation.

  \item[\textbf{R2.}] \textbf{Suggest Non-Obvious Connections:}
    Creativity is often found when a connection is formed between two incongruous objects or ideas~\cite{Kudrowitz2010improvisational,
    Ludden2012surprise}.
    A mind map can be considered as a network of connected ideas.
    A creative mind map can then be characterized as a network containing nodes such that at least some pairs of connected nodes have relationships between them that are not immediately obvious.
    This non-obviousness can be seen as an ambiguous representation, which allows reflection and reinterpretation of forms and relationships~\cite{Gross1996ambiguous}, in turn promoting creativity~\cite{Toh2016choosing}.
    An absence of explicit explanation for the relationship between connected nodes on a mind map can thus form the basis of reinterpretation.

  \item[\textbf{R3.}] \textbf{Self-Organize:}
    In our preliminary study, we noticed long,  convoluted links between concepts physically separated by distance and by other concepts that lay in between.
    These convoluted links make it difficult to perceive the connection between concepts~\cite{kobourov2015gestalt}.
    A digital mind mapping tool should re-organize so that connected concepts are close together.
    
  \item[\textbf{R4.}] \textbf{Allow Reorganization:}
    A function often found in existing mind-mapping tools is the freedom to reorganize concepts and connections, and creating new, non-hierarhical connections.
    This reorganization aids reflection and a better understanding of problems and solutions~\cite{Kokotovich2008problem}.
    It is thus important to provide the flexibility to change the layout of the mind map, and even the way in which nodes and links are connected.

\end{enumerate}

\vspace*{-7pt}

\subsection{Design and Implementation}
Based on the requirements identified above, and based on how one approaches mind mapping for creative problem solving, we decided on the following system design.

\textbf{Selecting a Knowledge Graph:}
In a mind-mapping process, the creator writes down the root idea (typically a representation of the problem at hand) and thinks of considerations, constraints, or sometimes serendipitous associations.
These associations emerge as part of the user's own knowledge and thought process.
Since \techname\ aims to emulate this process, it needs to be powered by an existing knowledge graph based on which associations can be suggested to the user.
Though there exist domain-specific knowledge repositories for concept generation (e.g.,~\cite{Bohm2008using}), these are intended for specific design processes and are not a substitute for the free and unusual associations that we intend to suggest through \techname.
We chose ConceptNet~\cite{Speer2017} as the knowledge graph to use as it is not domain-specific and represents general knowledge and concept associations.
The domain non-specificity can provide concept associations that have the potential to help the creator of the mind map go beyond familiar concept associations, which is one of the ways we prevent fixation (\textbf{R1}).

\textbf{Generating Suggestions:}
When a mind map creator wants to create new concept nodes that they associate with an existing node, they may have one of two inclinations.
For a very open-ended scenario where the goal is to explore novel ideas, they may create nodes whose associations with the existing node may not be obvious, akin to ``free association''~\cite{Freedman1965increasing}.
For more constrained problems, they may create nodes with more apparent relationships to the existing node.
To simulate these two approaches, we use a biased random walk to explore the nodes in the neighborhood of the knowledge graph described earlier.
We use the approach suggested by Grover et al.~\cite{Grover2016node2vec}, who create a search bias by defining a second-order random walk with two parameters.
The first is an in-out parameter $q$, such that a value of $q<1$ directs the walk away from the origin node, and $q>1$ directs the walk towards the starting node.
The second is a return parameter $p$ that determines, during a random walk, the likelihood of returning to an already-visited node in the graph.
A value of $p>max(q,1)$ biases the walk against visiting an already-visited node, while $p<min(q,1)$ backtracks the walk to keep it closer to the starting node.
Thus, we can bias the random walk towards a breadth-first search ($p<min(q, 1)$, $q>1$) that suggests nodes whose relationships with the origin node are apparent, or towards a depth-first search ($p>max(q,1)$, $q<1$) that suggests nodes whose relationships with the source node are not obvious (\textbf{R2}, see Fig.~\ref{fig:interface}).
As creative stimuli, distantly-related or unrelated text work better than image-based suggestions~\cite{Goncalves2012far}.
We thus provide text suggestions rather than images or other media.

\textbf{Withholding Explanations:}
One of the main differences between mind maps and concept maps
is that concept maps---typically used for understanding and explanation, rather than ideation---include annotated edges that describe the nature of the links between concepts~\cite{Eppler2006comparison}.
Mind maps do not typically describe each edge, nor does each edge in a mind map represent the same kind of relationship.
Prior work by Chen et al.~\cite{Chen2019mini} has illustrated the use of explanations as a way to prompt different aspects of the design
problem to consider, but we are interested in suggesting non-obvious associations for the user to make.
In providing suggestions through the biased random walk, and withholding the relationships between the source and suggested concepts, we intend to make the connections between nodes non-obvious, requiring reflection and reinterpretation from the user (\textbf{R2}).
We also decided to eschew directional links: while the topology of the mind map (center--out, mostly hierarchical) implies a directionality, we observed that participants did not use directionality meaningfully.

\textbf{Interactions:}
Observations from our preliminary study indicated that a flexibility in structure (e.g., forming non-hierarchical connections) and the ability to re-organize the mind map (especially when it grew to span multiple sheets) were clear requirements.
Not only is flexibility in reorganization a practical advantage offered by the digital medium, it also helps reflection and reinterpretation of concepts and relationships.
We thus adopt the convenience of easy editing and flexibility afforded by digital mind-mapping tools.
We use a force-directed layout approach to self-organize the map (\textbf{R3}) so that connected nodes stay closer to each other.
The user is still allowed to move and reorganize nodes, as well as delete existing links and add new ones (\textbf{R4}).
The suggestions are shown only when the user ``demands'' them, in keeping with findings by Siangliulue et al.~\cite{Siangliulue2015providing}, who recommend that suggestions generated on demand stimulate creativity better than suggestions provided at regular intervals.
Finally, keeping in mind that \techname\ is also meant to inspire new directions of thought, we facilitate manual node creation where the user can add their own ideas to the mind map.


\section{User Study}

\techname\ actively aids the mind-mapping process by ``suggesting'' non-obvious associations---concepts that are linked to the source concept, albeit indirectly.
These suggestions are meant to aid divergent thinking  by helping the user think of new concepts and relationships.
This delays fixation and allows them to consider associations that result in original ideas.

We have seen through prior studies on ideation and mind mapping processes that: (a) a greater number of nodes in a mind map is correlated with more creative ideas~\cite{Leeds2019mapping}, (b) a greater depth---distance from the central ``root'' node to the leaf node---is correlated with more creative ideas~\cite{Leeds2019mapping}, and (c) fluency (i.e.\ a greater quantity) and variety of ideas can be used as a measure of creativity~\cite{Shah2003}.
In evaluating \techname\ through a user study, our goal is to determine if participants created more nodes, a mind map with a greater average depth, and created more diverse concepts  when compared to a baseline mind-mapping tool that offered no suggestions.
We conduct a controlled study where participants were asked to create mind maps for two creative prompts: one for a specific application area and thus constrained, and the other more open ended. We use these two task categories to evaluate the impact of breadth-first and depth-first search biases.

\subsection{Participants}

We recruited 24 paid participants (14 female, 10 male), aged 18--34 years.
Ten of the participants were Ph.D.\ students, 4 were Master's students,
and 8 were undergraduate.
Of the student participants, 6 were design majors, 11 computer
science, 2 chemical engineering, 1 cognitive science, and 1 major in
international relations.
There were also three non-student participants, 2 with backgrounds in design and 1 in cognitive science.
On a scale of 1 (not familiar with mind maps)  to 7 (use them regularly), the users rated themselves a median of 5
(\begin{sparkline}{5}
\sparkspike .083 0.00
\sparkspike .226 0.09
\sparkspike .369 0.45
\sparkspike .512 0.45
\sparkspike .655 1
\sparkspike .797 0.27
\sparkspike .940 0.09
\end{sparkline})
on their familiarity with mind maps.

\subsection{Conditions and Tasks Design}
To study the effect of using Spinneret on mind mapping for open-ended and constrained tasks, we devised two tasks.
The constrained task was to think of solutions to reduce distracted driving.
The unconstrained task was to come up with a completely original movie plot.
In both cases, participants were to create a mind map that showed their thought process, including their considerations  when addressing the given problem.

Since we needed a baseline interface to compare with \techname, we created a modified version of \techname\ that provided the same organization and interaction capability, but without the suggestions.
We chose this approach instead of using existing mind-mapping tools to reduce confounding effects that may emerge from other aspects of the interfaces.

We thus constructed four task-interface combinations using the two tasks and the two interfaces.
Each participant was required to complete two task-interface combinations such that they used the baseline interface for one task and \techname\ for the other.
The study was counterbalanced to mitigate learning effects.
Thus, equal numbers of participants used \techname\ before and after the baseline, and equal numbers performed the constrained task before and after the unconstrained task.

\begin{figure}[b]
\vspace{-0.5em}
    \centering
    \includegraphics[width=0.8\columnwidth]{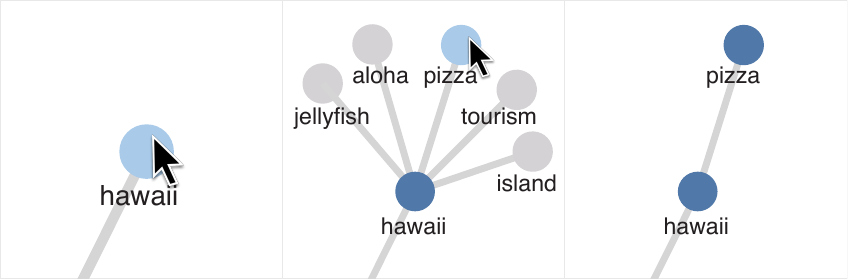}
    \caption{Users can click on a concept node to see
      \techname-suggested nodes for that concept. Depending on the
      parameters, the suggestions may result from a biased random walk
      with a depth-first or breadth-first search bias.
    In this case, the user clicks on ``hawaii'' and sees the suggestions \emph{jellyfish, aloha, pizza, tourism}, and \emph{island}.
    The user likes the concept ``pizza'' and accepts the suggestion by selecting the node.}
    \label{fig:spinneret-interactions}
\end{figure}

\subsection{Experimental Setup}
Both the \techname\ and baseline interfaces were run on a 2015 MacBook Pro (16 GB RAM, 2.8 GHz processor). 
Both interfaces were displayed on a 27-inch display (2560 $\times$ 1440 pixels), using the Google Chrome browser.
Participant input was provided through an external keyboard and mouse.

For the \techname\ interface, we added a few constraints.
In \techname, entering the root node in the text entry field 
adds not only the root node, but five neighbors (instead of suggestions)  to create an initial mind map.
This was done to provide prompts for the user to think in different directions from the start.
For subsequent nodes, the user needs to interact with any given node to display suggestions.
They then have two options: select one suggestion (dismissing the rest automatically), or dismiss all suggestions.
This prevented the user from being overwhelmed with what would otherwise be a geometric explosion of nodes in the mind map.

Another goal with the study is to determine what kind of suggestions work best, and for what task.
The two kinds of suggestions are those generated through a random walk with (1) a depth-first search bias and (2) a breadth-first search bias.
We thus set up \techname\ to randomly choose values of the in-out parameter ($q$) and the return parameter ($p$) such that the suggestions toggled between breadth-first and depth-first biased random walks every time the user requests suggestions.
We logged the suggestions and the $p$ and $q$ values for each user over the course of the tasks.

Finally, we used the largest component of the network of English concepts in ConceptNet 5.7.0 as the knowledge graph for suggestion. 
To ensure that every node can generate suggestions in the \techname\ condition (e.g., Fig.~\ref{fig:spinneret-interactions}), we constrained \techname.
Participants could manually add a node only if the text in the node matched text that existed in ConceptNet.
To ease this process, the manual text entry field (and the field for entering the main ``root'' node) both had autocomplete suggestions to help users select the text closest to what they intended.
In order to ensure that participant solutions are not constrained  by the text limitation, we provided participants with a pen and paper to jot down solutions for each task.
For the sake of uniformity, pen and paper were provided to participants regardless of the condition and task.

\subsection{Procedure}

Each participant first filled in their background and demographics in a survey form.
They were then trained on the first assigned interface (baseline/\techname) until they were comfortable using it.
They were then asked to generate a mind map to explore all considerations for the given task (constrained/unconstrained), and write down multiple solutions on the provided sheet of paper.
At the end of each task, participants responded to a questionnaire about their satisfaction with their solutions, and were asked to identify nodes on the mind map that were less relevant to their task.
They also responded to questions on the NASA TLX scale~\cite{Hart1988development} for each task. 
We also used a concurrent think-aloud protocol, and participants were audio- and screen-recorded for the duration of the tasks.
Participants' mind maps, user logs (including suggestions and search parameters), and list of ideas were all saved.

\section{Results}

\subsection{Quality of Mind Maps}

We measure indicators such as number of nodes, diversity of concepts used in the mind maps to characterize any differences between \emph{baseline} and \emph{\techname} used for the \emph{constrained} and \emph{unconstrained} tasks.

\subsubsection{Network Analysis}

Network measures such as the number of nodes and depth are known to be correlated to the quality of a mind map~\cite{Leeds2019mapping}.
For the \emph{unconstrained} task, Welch's t-test showed no significant difference ($p = 0.628$) in the number of nodes in the mind maps between \emph{\techname} ($M = 29.67$, $\sigma = 9.893$) and \emph{baseline} ($M = 32.33$, $\sigma = 15.93$).
Nor did we find a significant difference in the number of nodes between \emph{\techname} ($M = 26.58$, $\sigma = 10.30$) and \emph{baseline} ($M = 27.75$, $\sigma = 5.643$) for the \emph{constrained} task ($p = 0.735$).

Using Welch's t-test to compare the mean depths of the mind maps showed no significant difference ($p = 0.806$) between \emph{\techname} ($M = 2.945$, $\sigma = 1.908$) and \emph{baseline} ($M = 2.783$, $\sigma = 1.191$) for the \emph{unconstrained} task.
No significant difference was found in the mean depth between \emph{\techname} ($M = 2.241$, $\sigma = 0.747$) and \emph{baseline} ($M = 2.329$, $\sigma = 0.708$) for the \emph{constrained} task ($p = 0.770$) either.

\subsubsection{Diversity of Concepts}

The degree of non-obvious associations in the mind map can be measured using the notion that concepts different in meaning, context, and co-occurrence would not be easily associated with each other.
A word embedding technique allows individual words or phrases to be represented as vectors of real numbers, which can then measure semantic similarities between different words.
Hence, we use word embeddings to compute the association between any two concepts in the mind map using pairwise cosine distances.
We then measure the diversity of each mind map using the mean of the pairwise cosine distances of concepts in the mind map.
With ConceptNet as our knowledge graph, we use the word embedding computed with ConceptNet, i.e., ConceptNet Numberbatch~\cite{Speer2017}.
However, not every concept in the mind maps exists in ConceptNet Numberbatch.
In order to minimize the number of discarded concepts, we tokenized n-grams, lemmatized individual words, and fixed obvious typographical errors.
This resulted to discarding three concepts (\emph{Neuralink, Waze, \emph{and} Taika Waititi}) from the total 1,501 concepts as they do not exist in Numberbatch.
We also compute effect sizes using Cohen's $d$ with pooled standard deviation.

For the \emph{unconstrained} task, Welch's t-test showed that 
participants created significantly more diverse
mind maps  ($t(18.06) = 3.470$, $p = 0.003$, $95\%$ CI $= 0.013$--$0.053$) using \emph{\techname} ($M = 0.954$, $\sigma = 0.017$) than \emph{baseline} ($M = 0.922$, $\sigma = 0.028$), where the effect size is $1.416$ (large).
We found similar results for the \emph{constrained} task; mind maps were significantly more diverse  ($t(21.98) = 3.379$, $p = 0.003$, $95\%$ CI $= 0.013$--$0.055$)  when created with \emph{\techname}  ($M = 0.937$, $\sigma = 0.024$) than \emph{baseline} ($M = 0.902$, $\sigma = 0.025$), where 
the effect size is $1.380$ (large).

The suggested concepts were more diverse ($t(2943.3) = 10.77$, $p < 0.001$, $95\%$ CI $= 0.046$--$0.067$) with the DFS approach ($M = 0.949$, $\sigma = 0.112$) than the BFS approach ($M = 0.892$, $\sigma = 0.194$), where the effect size is $0.355$ (small).

\begin{figure}[t]
    \vspace{-1em}
    \centering
    \subfloat[diversity of mind map]{\includegraphics[]{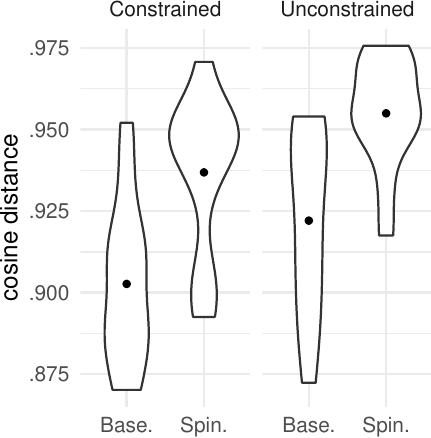}}\hfill
    \subfloat[distinctness of concept]{\includegraphics[]{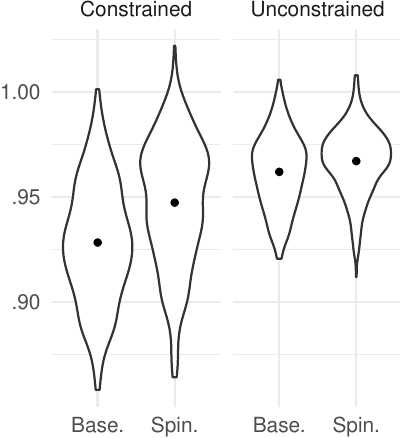}}
    \caption{Distributions of mind map diversity and concept distinctness 
      using the cosine distances between concept nodes as a metric.
    The point in each violin plot represents the mean.
    The distribution of diversity within mind maps clearly differs between baseline (Base.) and {\techname} (Spin.) for both tasks. 
    Based on the distribution of concept distinctness across all mind maps, we can see that participants created more distinct concepts for the unconstrained task than the constrained task.}
    \label{fig:result-diversity-uniqueness}
    \vspace{-1em}
\end{figure}

\subsubsection{Distinctness of Concepts}

We also compute how unique or distinct a concept is in comparison to other concepts generated in the study.
To do this,  we measure the mean of the cosine distances from one concept to all the other concepts from all the mind maps created by all the participants.
Thus, if the concept has greater distances from the other concepts than those concepts themselves, we believe that concept is more \textit{distinct} than the others.

For the \emph{unconstrained} task, Welch's t-test showed that the
participants came up with more distinct concepts ($t(762.5)
= 4.293$, $p = 1.991 \times 10^{-5}$, $95\%$ CI $= 0.003$--$0.007$)
using \emph{\techname} ($M = 0.967$, $\sigma = 0.016$) than
\emph{baseline} ($M = 0.962$, $\sigma = 0.018$),
where the effect size is 0.306 (small).
We also found the similar result for the \emph{constrained} task, the participants came up with more distinct concepts ($t(657.8) = 8.494$, $p < 2.2 \times 10^{-16}$, $95\%$ CI $= 0.015$--$0.023$) using \emph{\techname} ($M = 0.947$, $\sigma = 0.030$) than \emph{baseline} ($M = 0.928$, $\sigma = 0.029$), where the effect size is 0.646 (medium).

\begin{figure}[t]
    \centering
    \includegraphics[width=\columnwidth]{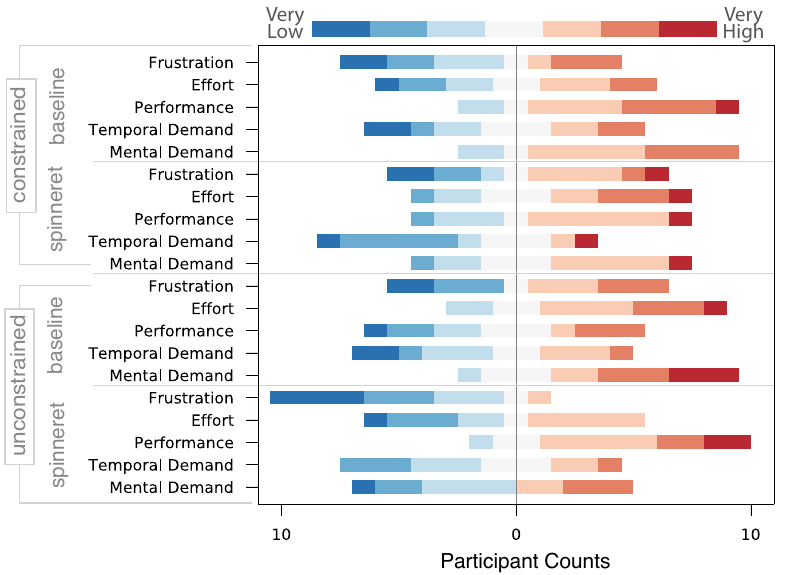}
    \caption{NASA TLX responses. Each stacked bar represents a group of twelve users who used one of the four combinations of two tasks (constrained and unconstrained) and two tools (baseline and \techname).}
    \label{fig:likert-nasa-tlx}
    \vspace{-1em}
\end{figure}

\subsection{Participant Feedback}

To compare the differences in participant feedback on the NASA TLX scale between \emph{tools}, we use the Brunner-Munzel test~\cite{BrunnerMunzner2000, Neubert2007}, which is a modification of the Wilcoxon-Mann-Whitney test. 
The Brunner-Munzel test is more robust for a test of medians than the Wilcoxon-Mann-Whitney test as it is not sensitive to differences in the shape of the distributions~\cite{Fagerland2009, Divine2018}.
Specifically, we use the permuted Brunner-Munzel test~\cite{Neubert2007} as it is robust even for a small sample size~\cite{Divine2018}.

For the \emph{unconstrained} task, the participants reported significantly less mental demand ($p = 0.022$) with \emph{\techname} ($Md = 3$, $IQR = 2.5$) than \emph{baseline} ($Md = 5.5$, $IQR = 2.25$).
For the \emph{unconstrained} task, the participants reported
significantly less effort  ($p = 0.021$) using \emph{\techname} ($Md = 3.5$, $IQR = 3$) than \emph{baseline} ($Md = 5$, $IQR = 2$).
There were no significant differences between \emph{tools} for other indices such as performance, frustration, and temporal demand.
We did not find any significant difference for the \emph{constrained} task (Fig.~\ref{fig:likert-nasa-tlx}).

\subsection{Ideas Generated in the Tasks}
Comparing the number of ideas between the two \emph{tools} using Welch's t-test showed that for the \emph{constrained} task, the participants created significantly more ideas ($t(19.71) = 2.13$, $p = 0.046$, $95\%$ CI $= 0.065$--$6.102$) using \emph{baseline} ($M = 9.5$, $\sigma = 4.1$) than \emph{\techname} ($M = 6.42$, $\sigma = 2.87$), where the effect size is 0.871 (large).
No significant difference was observed for the \emph{unconstrained} task.

For the \emph{unconstrained} task, the permuted Brunner-Munzel test showed a higher satisfaction (on a 7-point Likert) with solution exploration among participants ($p = 0.028$) using \emph{\techname} ($Md = 6$, $IQR = 0.25$) than \emph{baseline} ($Md = 4.5$, $IQR = 3.25$).
No significant difference in satisfaction was observed for the \emph{constrained} task.

For the \emph{unconstrained} task, the permuted Brunner-Munzel test revealed that the participants rated their ideas significantly higher ($p = 0.047$) using \emph{\techname} ($Md = 6$, $IQR = 1.25$) than \emph{baseline} ($Md = 4$, $IQR = 3$).
No significant difference in participant rating was observed for the \emph{constrained} task.

\subsection{\techname\ and Task Type}
Aside from comparing the mind-mapping experience and outcomes between \emph{baseline} and \emph{\techname}, we also are interested in
examining the suitability of \techname\ for the \emph{constrained} and \emph{unconstrained} tasks.
We thus performed a further analysis \emph{within} \techname\ on the suggestions that we report below.

\subsubsection{Accepted Suggestions}
A chi-square test showed a statistically significant association between task type and whether or not a suggestion was accepted ($\chi^2(1) = 16.81$, $\phi = 0.151$, $p = 4.138 \times 10^{-5}$),
where $45.1\%$ (183 of 406) of suggestions are accepted for \emph{unconstrained} task whereas  $30.3\%$ (101 of 333) of suggestions are accepted for \emph{constrained} task.

We did not find any differences of the acceptance of suggestions between the type of biased random walks (i.e.\ breadth-first search and depth-first search), nor the parameters governing the bias---$p$ and $q$---for both tasks.

To understand the difference of the accepted suggestions between the two tasks, we analyzed the word embedding distance between a suggestion and its corresponding source concept.
Welch's t-test showed that the accepted suggestions are significantly farther ($t(177.2) = 2.182$, $p = 0.030$, $95\%$ CI $= 0.005$--$0.094$) from their source concepts in the \emph{unconstrained} task ($M = 0.890$, $\sigma = 0.161$) than the \emph{constrained} task ($M = 0.840$, $\sigma = 0.193$), where the effect size is 0.285 (small).

\subsubsection{Overall Feedback}

Finally, participants reported their overall feedback about \emph{\techname} on a 5-point Likert scale.
Participants who used \emph{\techname} for the \emph{unconstrained} task rated \emph{\techname} significantly higher ($Md = 3.5$, $IQR = 1$, 
\begin{sparkline}{3}
\sparkspike .083 0.0
\sparkspike .283 0.33
\sparkspike .483 0.67
\sparkspike .683 1.00
\sparkspike .883 0
\end{sparkline}
) for its relevance to their task ($p = 0.005$) than the participants who used \emph{\techname} for the \emph{constrained} task ($Md = 2$, $IQR = 0.25$,
\begin{sparkline}{3}
\sparkspike .083 0.13
\sparkspike .283 1.00
\sparkspike .483 0.25
\sparkspike .683 0.13
\sparkspike .883 0
\end{sparkline}
).

Most participants reported that \emph{\techname} often suggested options that made them think of ideas they would not have come up with otherwise for both \emph{unconstrained} ($Md = 5$, $IQR = 1$, 
\begin{sparkline}{3}
\sparkspike .083 0.00
\sparkspike .283 0.00
\sparkspike .483 0.29
\sparkspike .683 0.43
\sparkspike .883 1.00
\end{sparkline}
) and \emph{constrained} ($Md = 4$, $IQR = 0.25$,
\begin{sparkline}{3}
\sparkspike .083 0.00
\sparkspike .283 0.00
\sparkspike .483 0.29
\sparkspike .683 1.00
\sparkspike .883 0.57
\end{sparkline}
) tasks, but the differences are not statistically significant.

While the participants who used \emph{\techname} for \emph{unconstrained} reported that the suggestions are less often confusing ($Md = 3.5$, $IQR = 2$,
\begin{sparkline}{3}
\sparkspike .083 0.17
\sparkspike .283 0.50
\sparkspike .483 0.33
\sparkspike .683 1.00
\sparkspike .883 0.00
\end{sparkline}
) than for \emph{constrained} (median~$=4$, IQR~$=1$,
\begin{sparkline}{3}
\sparkspike .083 0.00
\sparkspike .283 0.33
\sparkspike .483 0.33
\sparkspike .683 1.00
\sparkspike .883 0.50
\end{sparkline}
), the difference are not statistically significant.

While the participants who used \emph{\techname} for \emph{unconstrained} reported that they are more likely to use ($Md =4$, $IQR =0.25$,
\begin{sparkline}{3}
\sparkspike .083 0.00
\sparkspike .283 0.29
\sparkspike .483 0.14
\sparkspike .683 1.00
\sparkspike .883 0.29
\end{sparkline}
) than for \emph{constrained} ($Md =3,5$, $IQR =1.25$,
\begin{sparkline}{3}
\sparkspike .083 0.00
\sparkspike .283 0.60
\sparkspike .483 0.60
\sparkspike .683 1.00
\sparkspike .883 0.00
\end{sparkline}
), the differences are not statistically significant.

\begin{figure}[t]
    \centering
    \includegraphics[width=\columnwidth,keepaspectratio]{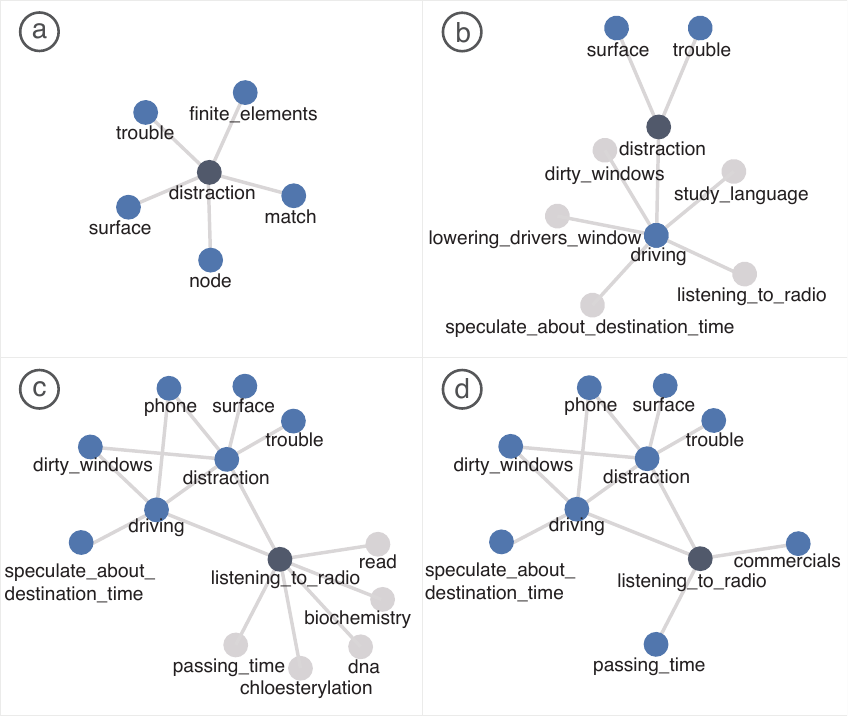}
    \caption{A walk-through of the general behavior of users who use Spinneret for the constrained task.}
    \label{fig:spinneret_car}
    \vspace{-1em}
\end{figure}

\section{Task Characterization}

The results show that users are more likely to select from the suggestions offered by \techname\ in unconstrained tasks.
Based on our observation and review of participants' screen recording accompanied by their think-aloud, we describe the behavior of two participants when using \techname\ for a typical open-ended task and a typical unconstrained task.

\subsection{Seeking (specific) Inspiration: Driving Distractions}

In the case of the constrained task, i.e.,\ ``find ways to reduce distractions when driving'', the context and scope of the task is somewhat limited.
Let us consider a participant, Alice, who is given this task to execute with the \techname\ interface.
Alice represents the general behavior for users using \techname\ for the constrained task: she has an idea of what she wants to find among the suggestions due to the task constraints.
She uses \techname\ mainly as a way to inspire alternative ways to explore the design space and finds herself using the suggestions only when she runs out of ideas.

Alice is an experienced driver and has a good idea of driving distractions.
She starts her mind map by typing in ``distraction'' in the main input box (Fig.~\ref{fig:spinneret_car}a).
Among the suggestions, she sees \emph{trouble}, \emph{finite elements}, \emph{surface}, \emph{node}, and \emph{match}.
\emph{Surface} reminds Alice of driving on different surfaces, so she keeps that node.
She finds \emph{finite element} and \emph{match} irrelevant, so she deletes them.
Since she does not see anything relevant to driving, she manually creates a node called \emph{driving}.
She then pulls up suggestions related to \emph{driving} (Fig.~\ref{fig:spinneret_car}b), and sees the terms \emph{speculate about destination time, listening to radio, lowering driver's window, dirty window}, and \emph{study language} come up.
She is pleased with the suggestions---some of them are what she already had in mind---and proceeds to accept \emph{lowering driver's window, speculate about destination time, listening to radio} and \emph{dirty window}, and then manually adds \emph{phone} alongside the new nodes on the mind map.

Looking at the \emph{listening to radio} node, Alice wonders what might help people when they are distracted by the car radio.
She brings up the suggestions (\emph{noise, textile, pony, reading magazine}, and \emph{factory}), but does not find them relevant.
She tries again and gets \emph{biochemistry, dna, passing time, cholesterylation}, and \emph{read} (Fig.~\ref{fig:spinneret_car}c).
She agrees that \emph{listening to the radio} is a way of \emph{passing time} and picks that node.
Alice recalls her own experience of being distracted with the radio whenever a commercial break occurs, as she would try to change the radio to a different station.
So she adds in the \emph{commercial} node and comes to the final concept of ``Research different \textbf{radio stations' commercial} breaks'' and ``have timely calendar reminders to avoid \textbf{dirty windows}" (Fig.~\ref{fig:spinneret_car}d).

\subsection{Free Association: Original Movie Plot}

``A completely original plot for a movie'' represents an instance of blue-sky thinking: creative thinking that is not limited by prevalent norms and constraints.
In this situation, concepts and associations can be incongruous without being irrelevant.
Let us consider a typical participant, Bob, who uses \techname\ for this task.
Bob is aware that an original idea can stem from trying to create relationships between concepts seem incongruous at first.

\begin{figure}[h]
    \centering
    \includegraphics[width=\columnwidth]{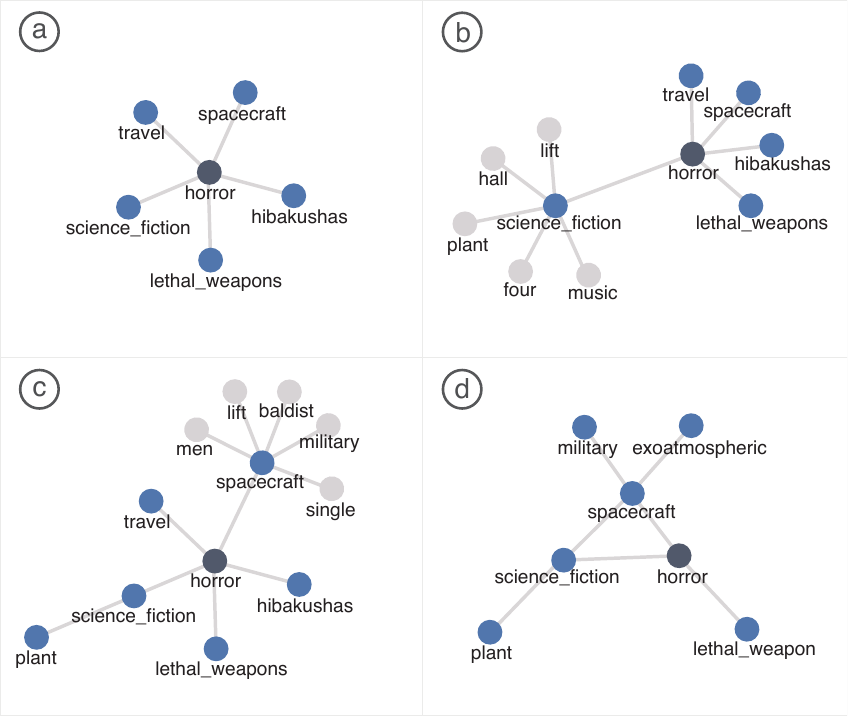}
    \caption{A walk-through of the general behavior of users who use Spinneret for the unconstrained task.}
    \label{fig:spinneret-movie-frame}
    \vspace{-1em}
\end{figure}

Bob starts by typing in his favorite genre---horror---as the root node.
He immediately sees the motley set of suggestions offered: \emph{hibakushas}, \emph{lethal weapons}, \emph{travel}, \emph{spacecraft}, and \emph{science fiction} (Fig.~\ref{fig:spinneret-movie-frame}a).
He is intrigued by the possibility of combining science fiction and horror, and brings up suggestions related to the \emph{science fiction} node.
He sees the suggestions \emph{lift, music, hall, plant, and four} (Fig.~\ref{fig:spinneret-movie-frame}b).
He decides his sci-fi horror movie will involve plants, and selects the corresponding node.
Looking back at the root node, the node \emph{spacecraft} attached to it catches his eye.
He brings up suggestions for \emph{spacecraft}, and finds the word \emph{military} interesting (Fig.~\ref{fig:spinneret-movie-frame}c).
This process continues and he devises a movie plot of  ``An \textbf{exo-atmospheric} \textbf{military spacecraft} tries to kill a \textbf{horror} \textbf{plant} with a \textbf{lethal weapon}'' (Fig.~\ref{fig:spinneret-movie-frame}d). 

\section{Discussion}

\subsection{Distinctness of Concepts}
An analysis of the mind map showed that there was no significant difference in participant fluency  (measured in number of nodes created), or in the average mind-map depth between the baseline and \techname\ conditions for both tasks. 
Yet, concept node distinctness among participants using \techname\ was significantly higher than participants using the baseline.
This contrasts with findings on mind-mapping exercises studied by Leeds et al.~\cite{Leeds2019mapping} who found that in order to generate more distinct concept nodes, one needs to create a higher number of nodes, and create mind maps with a greater depth.
This difference may be due to \techname\ providing non-obvious suggestions early.
This prompts the user to explore more unusual ideas that may otherwise take longer to reach, or not at all.
This is supported by 83\% of the participants (
\begin{sparkline}{4}
\sparkspike .083 0.00
\sparkspike .283 0.00
\sparkspike .483 0.40
\sparkspike .683 1.00
\sparkspike .883 1.00
\end{sparkline}
) who agreed or strongly agreed, regardless of the task for which they used \techname, that the tool prompted them to think of ideas they would not have otherwise considered.
This indicates that \techname\ does indeed help reduce design fixation (\textbf{R1}).

To verify this claim, we computed the correlation between the mean distinctness and the number of nodes only for the baseline interface's mind maps.
However, we failed to find the correlation that Leeds et al.~\cite{Leeds2019mapping} found.
This may be because Leeds et al.\ use manual coding to determine categorically whether two concepts mean the same (or not), while we use cosine distances as a measure of relatedness.
This implies that our notion of ``distinctness'' is a more fuzzy---and
possibly robust---concept, compared to their categorical notion.

\subsection{Diversity of the Mind Map}
Mind maps created using \techname\ were significantly more diverse than the baseline.
This indicates that on average, mind maps created using \techname\ have concepts that are less related to each other.
It would thus be more difficult to characterize the relationship between two nodes in a mind map generated by \techname, suggesting that it fosters non-obvious connections~\cite{Kudrowitz2013does, Ludden2012surprise} and reduces fixation~\cite{Duncker1945problem, Jansson1991design} (\textbf{R2}).

\subsection{The Role of the Task}
We studied the use of \techname\ on constrained and unconstrained tasks to determine whether the kinds of non-obvious suggestions are perceived differently for different task types.

\subsubsection{Fluency of Ideas}
While there was no significant difference in the fluency of concept node creation, participants generated significantly fewer ideas using \techname\ when compared to the baseline interface for the constrained task.
This is partly because of the proportion of suggestions that participants ended up dismissing, which was significantly higher for the constrained task ($69.7\%$) than for the unconstrained task ($54.9\%$).
Observations showed that participants tried multiple unsuccessful attempts to find a suggestion that was more relevant to their task, which is difficult to produce when the task is specific.
As one participant reported when using \techname\ for the constrained task, \emph{``Sometimes the associations seemed so irrelevant that I didn't want to put much effort into thinking about them.''}, while another entreated \techname, \emph{``give me something useful, please!''}
This indicates that there may be a case for balancing the more apparent (i.e.\ more ``obvious'' suggestions) with the non-obvious ones, to suit a greater range of tasks.

\subsubsection{Diversity and Distinctness}

A closer look at the concept distinctness plots (Fig.~\ref{fig:result-diversity-uniqueness}) shows that the mean cosine distance for the unconstrained tasks is significantly greater than for the constrained task.
One reason for the distinctness of nodes across participants could be the nature of the task itself: there is no clear context common to movies that would persist among participants, but ``reducing distractions when driving'' provides certain driving-related contexts that would span multiple participants.
For unconstrained tasks, this finding aligns with prior studies that show that text-based inspiration is useful even if distantly related~\cite{Goncalves2012far}, and provided on demand~\cite{Siangliulue2015providing}.
One participant, when looking at suggestions for ``spook'' on the unconstrained task, was surprised to see ``programming'' and ``kill'' among some of the suggestions.
\textit{``I like how those two words were not what I thought of when I think of ``spook,''} she noted, eventually coming up with a plot of a German programmer set during Halloween.

\begin{figure}[t]
\vspace{-0.5em}
    \centering
    \includegraphics[width=0.9\columnwidth]{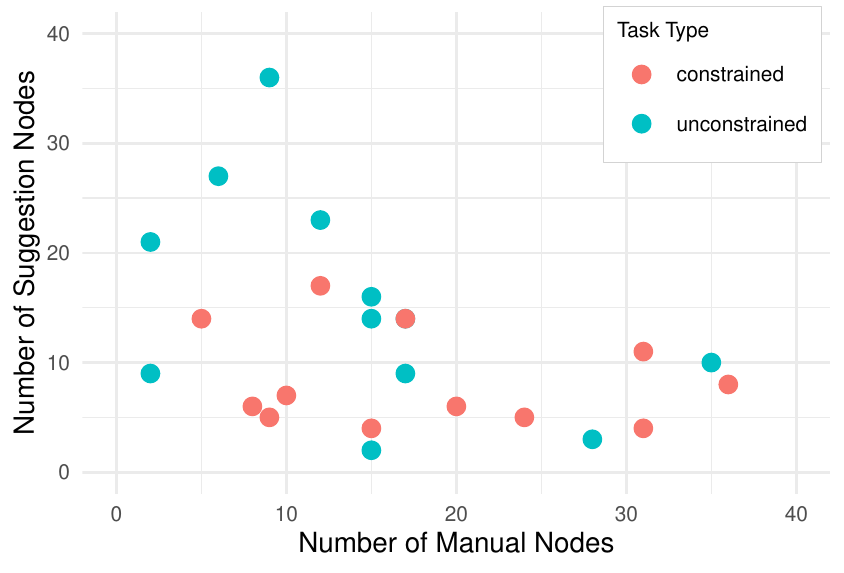}
    \caption{A scatterplot representing the mind maps created using \techname. Each point represents a mind map, colored by task type, with its x-position indicating the number of manually-created nodes and y-position indicating the number of \techname-suggested nodes in it.
    The scatterplot shows that mind maps created for the unconstrained task tended to have a greater proportion of \techname-suggested nodes.}
    \label{fig:spinneret-scatter}
    \vspace{-1em}
\end{figure}

Kudrowitz and Wallace~\cite{Kudrowitz2010improvisational, Ludden2012surprise} have suggested that incongruity can be seen as one of the drivers of creativity, which suggests that \techname\ would be more suitable for unconstrained tasks than constrained tasks.
As one participant mentioned that for the unconstrained task, \techname\ was \emph{``Almost like a game...what new suggestions can I get to pop up?''}
He noted that he found himself waiting for suggestions \emph{``to inspire the next move. If they're too related, then I'll think `no, not that'.''}
This is further brought into sharp relief in Fig.~\ref{fig:spinneret-scatter}.
While all mind maps in the chart represent those created using \techname, the ones for the constrained task have a consistently higher proportion of manual nodes than suggested ones.

\section{Limitations}
Our controlled study helped us understand the the influence of suggestions generated through a biased random walk on the distinctness and diversity of concepts in a mind map.
One of the ways in which we controlled the study was to limit the expressivity of the mind map, i.e.\ users could not change the size, color, or weight of nodes or edges.
Such expressivity is allowed in conventional mind mapping applications, which may allow users other ways in which to distinguish and organize concepts.
We plan to incorporate these features in future longitudinal studies.

The results show no significant difference in fluency (number of nodes) between baseline and \techname.
While this may indicate that \techname\ has no influence on idea saturation, we note that each task was limited to 15 minutes.
Longer, multi-session longitudinal studies will be needed to further investigate \techname's  effect on fluency.

Finally, one might argue that the comparison between \techname, which provides more ``information'' (as suggestions) and baseline, which does not, is unfair.
This is partly due to the nature of the study itself: our goal was to see if providing such information positively influences the outcome.
We also find that more information is not always better: participants found \techname's suggestions more useful for unconstrained tasks than for constrained tasks.
An analysis of cosine distances between the source node and suggested nodes showed that the cosine distances for accepted nodes in the constrained task tended to be smaller than the rejected nodes.
This difference in cosine distances between accepted and rejected nodes was lower for the unconstrained task.
There was no effect of the search bias (depth-first and breadth-first) on this outcome.
Simply providing more information is thus not enough: a better control of relevance is needed.
In the future, we are interested in exploring the use of word embeddings in place of a knowledge graph.
This may create new opportunities for controlling the relevance and obviousness of the suggestions and help generate useful mind maps.

\section{Conclusion}
In this paper, we present \techname, a tool that aids creative mind mapping by providing non-obvious suggestions to selected concepts.
We use a biased random walk with depth-first search and breadth-first search bias{es} on a knowledge graph to generate suggestions.
We have conducted a controlled study comparing \techname\ against a baseline interface for two mind-mapping tasks: a constrained task in the context of ``reducing distracted driving'' and an unconstrained task on generating an ``original movie plot''.
We find that mind maps generated using \techname\ showed a greater diversity in concepts, and concept nodes generated using \techname\ showed a higher uniqueness across participants.
We also find that \techname's suggestions were better received by participants for the unconstrained task than for the constrained task, regardless of the search bias applied.
We discuss implications for generating suggestions using word embeddings instead of knowledge graphs for a greater contextual control over the suggestions.

\section*{Acknowledgments}
We are grateful to Takanori Fujiwara and the anonymous reviewers for their thoughtful feedback that improved the content and readability of this work.
This research is sponsored in part by the US National Science Foundation through grants IIS-1741536 and IIS-1528203.


\balance{}

\bibliographystyle{SIGCHI-Reference-Format}
\bibliography{spinneret}

\end{document}